 \documentclass[final,3p,times,twocolumn]{elsarticle}

 \usepackage{graphicx}
 \usepackage{epstopdf}

\usepackage{amssymb}
\usepackage{amsmath}

\journal{Optics Communications}

\begin{document}

\begin{frontmatter}



\title{Stability analysis of polarization attraction in optical fibers}


\author[MIT]{Konstantin Turitsyn}
 \author[Brescia]{Stefan Wabnitz}
 \ead{stefano.wabnitz@ing.unibs.it} 
 
\address[MIT]{Department of Mechanical Engineering, Massachusetts Institute of Technology, Cambridge, 02139}
\address[Brescia]{Dipartimento di Ingegneria dell'Informazione,  Universit\`a degli Studi di Brescia, via Branze 38, 25123, Brescia, Italy}

\begin{abstract}
The nonlinear cross-polarization interaction among two intense counterpropagating beams in a span of lossless randomly birefringent telecom optical fiber may lead to the attraction an initially polarization scrambled signal towards wave with a well-defined state of polarization at the fiber output. By exploiting exact analytical solutions of the nonlinear polarization coupling process we carry out a linear stability study which reveals that temporally stable stationary solutions are only obtained whenever the output signal polarization is nearly orthogonal to the input pump polarization. Moreover, we predict that polarization attraction is acting in full strength whenever equally intense signal and pump waves are used.  

\end{abstract}


\end{frontmatter}


\section{Introduction}
\label{sec:intro}

The cross-interaction among intense counter-propagating beams in a Kerr or cubic nonlinear medium leads to a mutual rotation of their state of polarization (SOP). This effect was extensively theoretically studied since back in the 1980's: for example, Kaplan and Law \cite{kap85} found exact analytical solutions which exhibit polarization bistability and multistability, as later experimentally confirmed by Gauthier et al \cite{gaeta2}. The same process is also responsible for leading to both spatial \cite{yumo85}-\cite{swab86} as well as temporal \cite{gaeta1} polarization instabilities and chaos. The general spatio-temporal stability of the nonlinear polarization eigenarrangements (or eigenpolarizations) which remain unchanged upon propagation in the Kerr medium was analysed by Zakharov and Mikhailov \cite{zakh87}, who pointed out the formal analogy between the equations describing the Stokes vectors of the two beams and those associated with the coupling of spin waves in ferromagnetic materials or Landau-Lifshitz model. This led to the prediction of the formation of stable domains of mutual arrangements of SOPs which, depending upon the boundary conditions, may also produce all-optical polarization switching phenomena as experimentally observed with optical fibers by Pitois et al.\cite{pitois98}-\cite{JOSAB}.

The same nonlinear SOP interaction process also leads to an intriguing phenomenon known as polarization attraction, which has not been unveiled until relatively recently \cite{pitois04}-\cite{FatomeOE}. As a typical example of this effect, let us consider the case of a backward pump wave which is injected at one end of the Kerr medium with a given SOP. Then one may observe that the forward or signal beam, which is launched at the other end of the medium, emerges with a well-defined SOP, irrespective of its initial SOP. Thus we may say that the polarization interaction has led to the effective attraction of the output signal SOP towards a particular value which is determined by the SOP of the pump. The demonstration of such effect using CW beams in relatively long telecommunication fiber spans \cite{FatomeOE} paves the way for conceiving a new class of practical devices for the all-optical control of the signal SOP in optical communication and laser systems.

In recent years, a relatively large number of theoretical studies have permitted to derive the equations describing the SOP interactions of counter-propagating waves in randomly birefringent telecom fibers \cite{kozlov11a}-\cite{guaso_josab}. On the basis of these equations, it has been possible to reproduce the experimental findings with a good quantitative accuracy \cite{kozlov11}. From the analytical side, the study of the stationary (or time-independent) solutions has led to the interesting observation that polarization attraction is closely linked with the existence of singular tori or multi-dimensional separatrix solutions \cite{Asse11}-\cite{Asse12b}. Yet a rigorous analytical description of the process of relaxation of the signal wave SOP towards the attracting value remains largely elusive to date, with the exception of some relatively simple limit cases such as the wave reflection which occurs in a purely linear distributed feedback mirror \cite{Grenier}. Indeed, the main difficulty in the analysis of the problem is associated with the presence of boundary conditions in a medium of finite length.

In this work we present an advance in the understanding of the physical origin of polarization attraction, by showing that this effect is associated with the presence of spatio-temporally stable stationary solutions, whereas all other stationary solutions are unstable so that their decay towards the stable or attracting polarization arrangements is to be expected in the experiments. We consider the conservative polarization interaction between intense signal and counterpropagating pump beams in a randomly birefringent telecom optical fiber span. First we will derive a relatively simple closed-form expression linking the degree of relative polarization alignment between the pump and the signal at the output end of the fiber to their relative orientation at the fiber input. Next we will carry out a semi-analytical study of these solutions, showing that the only stable branch of solutions corresponds to the situation where the signal SOP is effectively orthogonal to the pump SOP as the fiber length (or beams' power) grows larger. 

Note that recent experiments have also unexpectedly revealed the effect of self-polarization, whereby a single beam interacting with its replica back-reflected at the fiber output end by a mirror also leads to the attraction towards circular polarization states, independently of the input SOP orientation \cite{Fatome}. A numerical study of the spatio-temporal stability of the stationary solutions has also permitted to associate the presence of attracting SOPs with the existence of stable branches \cite{Asse13}.    
   
\section{Basic Equations}
\label{sec:equations}

In this work we consider the polarization interaction of two intense CW beams counter-propagating
along the $z$ axis in a randomly birefringent fiber of length $L$. The evolution equations for the Stokes vectors of the forward (or signal) and backward (or pump) beams, ${\textbf S}^+=(S_1^+,\, S_2^+,\, S_3^+)^T$
and ${\textbf S}^-=(S_1^-,\, S_2^-,\, S_3^-)^T$ read, in dimensionless units, as \cite{kozlov_josab}
\begin{eqnarray}
\label{pde1}
&	\partial_\xi \textbf{S}^+ = \textbf{S}^+ \times \hat{J}_x \textbf{S}^- \nonumber \\
 \\
&	\partial_\eta \textbf{S}^-= \textbf{S}^- \times \hat{J}_x \textbf{S}^+ \nonumber
\end{eqnarray}
\noindent with distance $z = \xi - \eta$, time $t = \xi + \eta$; moreover $\times$ denotes vector cross product, and the cross-polarization tensor $\hat{J}_x = \mathrm{diag}\{-1,1,-1\}$. In the problem described by Eqs.(\ref{pde1}), two important physical parameters are the nonlinear length $L_{NL}\equiv 1/(\gamma S_0^+)$, where $\gamma$ is the nonlinear fiber coefficient, and the diffusion length of the polarization mode dispersion (PMD) $L_d^{-1}=\frac{1}{3}D_p^2(\omega_+-\omega_-)^2$. Here the PMD coefficient reads as

\begin{equation}
\label{dp}
D_p= \frac{2\sqrt{2}\pi\sqrt{L_c}}{L_B(\omega_+)\omega_+},
\end{equation}

\noindent where $L_c$ is the correlation length of the random birefringence process, $\Delta\beta (\omega_+)$ , and $L_B(\omega_s)=2\pi /\Delta\beta (\omega_s)$ are the linear birefringence and 
the beat length at the signal frequency $\omega_+$. With the aid of these definitions, it can be shown that Eqs.(\ref{pde1}) are valid in the so-called Manakov limit \cite{guaso_josab}, namely whenever $L,\, L_{NL}\ll L_d$.
Equations (\ref{pde1}) can be turned to symmetric form after switching to the new variables $\textbf{S} = \textbf{S}^+$ and $\textbf{H} = -\hat{J}_x \textbf{S}^-$ (or $\textbf{S}^- = -\hat{J}_x \textbf{H}$). The additional factor $\hat{J}_x$ in the definition of $\textbf{H}$ reflects the fact that the circular polarization is flipped for beams propagating in opposite directions.
\begin{align}\label{main1}
&(\partial_t + \partial_z) \textbf{S} = \textbf{H}\times \textbf{S} \\
&(\partial_t - \partial_z) \textbf{H} = \textbf{H}\times \textbf{S}  \label{main2}
\end{align}
The case of counter-propagating beams corresponds to the boundary conditions $\textbf{S}(0) = \textbf{S}_0$ and $\textbf{H}(L) = \textbf{H}_L$. In the following we will refer to the $\textbf{S}$ beam as the signal, and $\textbf{H}$ beam as the pump. We will assume that the input signal can have arbitrary polarization $\textbf{S}(0)$, that we will later assume to be uniformly distributed on the Poincare sphere. On the other hand the pump can have a controllable polarization $\textbf{H}(L)$. One of the primary goals of our study is to characterize the effect of polarization attraction, or in other words characterize the output polarization of the beam $\textbf{S}_L = \textbf{S}(L)$ and its relation to the controllable pump polarization $\textbf{H}(L)$. We will focus on the analysis of stationary solutions of (\ref{main1},\ref{main2}).

\section{Exact Solution}
\label{sec:solution}

The stationary solutions of (\ref{main1},\ref{main2}) can be found by noticing that whenever $\partial_t \textbf{S} = \partial_t \textbf{H} = 0$, the quantities ${\bf\Omega} = \textbf{H} +\textbf{S}$, as well as $H^2 = \textbf{H}\cdot\textbf{H}$ and $S^2 = \textbf{S}\cdot\textbf{S}$ will remain invariant throughout the fiber, i.e. $\partial_z {\bf\Omega} = \partial_z S = \partial_z H = 0$. The full solution of equations (\ref{main1},\ref{main2}) corresponds to a simple precession of the signal and the pump polarization vectors around the spatially constant vector ${\bf\Omega}$, therefore it can be written as: 

\begin{eqnarray}\label{SLdyn}
\displaystyle \textbf{S}(z) = \left(\textbf{S}(z') - \frac{{\bf \Omega}\cdot\textbf{S}(z')}{\Omega^2}\bf{\Omega}\right) \cos\Omega l +\nonumber \\
\displaystyle \frac{{\bf\Omega}\cdot\textbf{S}(z')}{\Omega^2}{\bf\Omega} +  \frac{\bf{\Omega}\times \textbf{S}(z')}{\Omega}\sin\Omega l
\end{eqnarray}
The same kind of relation also holds for the pump vector $\textbf{H}(z)$. In order to find the actual stationary solutions one needs to satisfy the boundary conditions, $\textbf{S}(0) = \textbf{S}_0$ and $\textbf{H}(L) = \textbf{H}_L$. We will do so by introducing two quantities that characterize the strength of polarization attraction in the fiber. First, we define the output signal-pump alignment factor
\begin{equation}\label{alig}
	\eta = \frac{(\textbf{H}_L\cdot \textbf{S}_L)}{H S}
\end{equation}
which measures the relative orientation of the signal and pump beams at the $z=L$ end of the fiber. The alignment factor (\ref{alig}) is the quantity that characterizes the strength of the polarization attraction effect: $\eta = 1$ corresponds to the situation where the signal SOP is the same as the pump SOP; whereas $\eta = -1$ corresponds to the situation where the signal and pump SOPs are orthogonal. 

However, given that the boundary conditions fix the polarization of the signal $\textbf{S}(0) = \textbf{S}_0$ at the left or $z=0$ boundary of the fiber, in order to find the value of $\eta$ we have to solve Eq.(\ref{SLdyn}) by assuming that the input signal-pump polarization alignment parameter 
\begin{equation}
	\mu = \frac{(\textbf{H}_L\cdot \textbf{S}_0)}{H S}
\end{equation}
is given. Note that a uniform initial distribution of the signal polarization on the Poincar\'e sphere (such as it is obtained from a SOP scrambler) corresponds to a uniform distribution of the scalar value $\mu$ on the interval $\mu \in [-1,1]$.

The relation between the output signal-pump polarization alignment $\eta$ and the corresponding intial alignment parameter $\mu$ can be found by using expression (\ref{SLdyn}) with $z = 0$ and $z' = L$. By assuming that the value of $\eta$ is given, so that the signal SOP the $z=L$ end of the fiber is fixed, we may find the initial signal-pump polarization alignment $\mu$ at the $z=0$ end of the fiber by taking the dot product of both sides of Eq.(\ref{SLdyn}) with $\textbf{H}_L$. This yields the following relation 
\begin{eqnarray}\label{Ffunc}
\displaystyle	\mu = F(\eta) = \frac{(H\eta + S)(H + S\eta)}{H^2 + S^2 + 2 H S \eta}\left(1-\cos\Omega L\right) + \nonumber \\
\displaystyle	 \eta \cos\Omega L
\end{eqnarray}
where we have used the condition $\Omega^2 =(\textbf{H} + \textbf{S})\cdot(\textbf{H} + \textbf{S})  = H^2 + S^2 + 2 H S \eta$. The desired relation between the output $\eta$ and initial $\mu$ polarization alignment parameters can be found by inverting the function $F(\eta)$, i.e., by solving the equation $\mu = F(\eta)$ for $\eta$.

\begin{figure}[ht]
\centering
\includegraphics[width=\columnwidth]{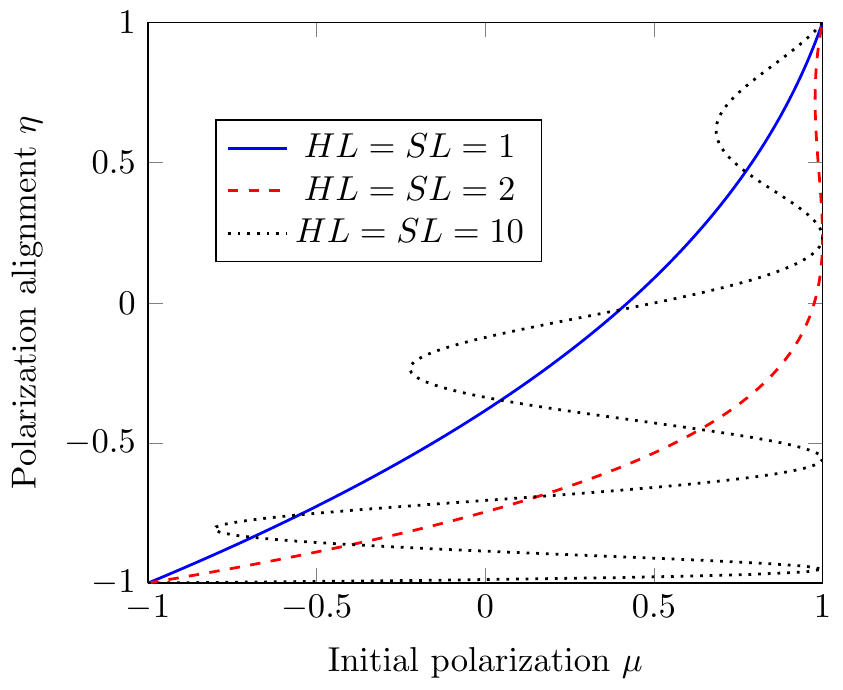}
\caption{Relation between output $\eta$ and input $\mu$ polarization alignment parameters, showing multiple branches of stationary solutions which are observed for high enough signal and pump powers $S L$ and $H L$.}
\label{fig-branches}
\end{figure}

\begin{figure}[ht]
\centering
\includegraphics[width=\columnwidth]{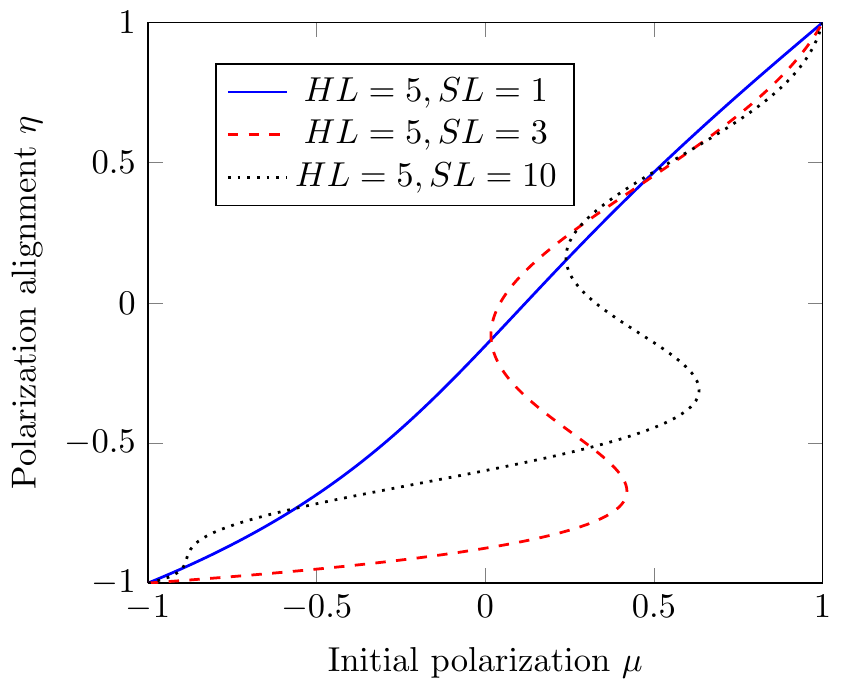}
\caption{Multiple branches of stationary solutions for non-matched values of  $S$ and $H$.}
\label{fig-new}
\end{figure}

Significant insight into the origin of the polarization attraction effect can be gained via a simple analysis of the algebraic curves which are defined by the relation (\ref{Ffunc}). As one can see from figure (\ref{fig-branches}), in general there are multiple branches of stationary solutions, that is there are potentially many stationary solutions corresponding to one and the same given value of the initial signal-pump polarization alignment $\mu$. In order to understand what kind of solution will be observed in practice, one needs to analyze the temporal stability of all of these solutions.  As will be shown in the next section, a numerical stability analysis shows that only the lowest branch of the solutions shown in Fig.(\ref{fig-branches}), corresponding to the smallest value of $\eta$, is stable. In other words, the output signal beam tends to get attracted to an SOP which is orthogonal to the pump. This observation is fully consistent with the results of the numerical analysis performed in ref. \cite{kozlov_josab}. Moreover, direct numerical simulations of the stationary solutions as those reported in Fig.(\ref{fig-branches} have confirmed that, in the case of multiple solutions, only the lowest branch is temporally stable \cite{kozlov_unpub}.

\begin{figure}[ht]
\centering
\includegraphics[width=\columnwidth]{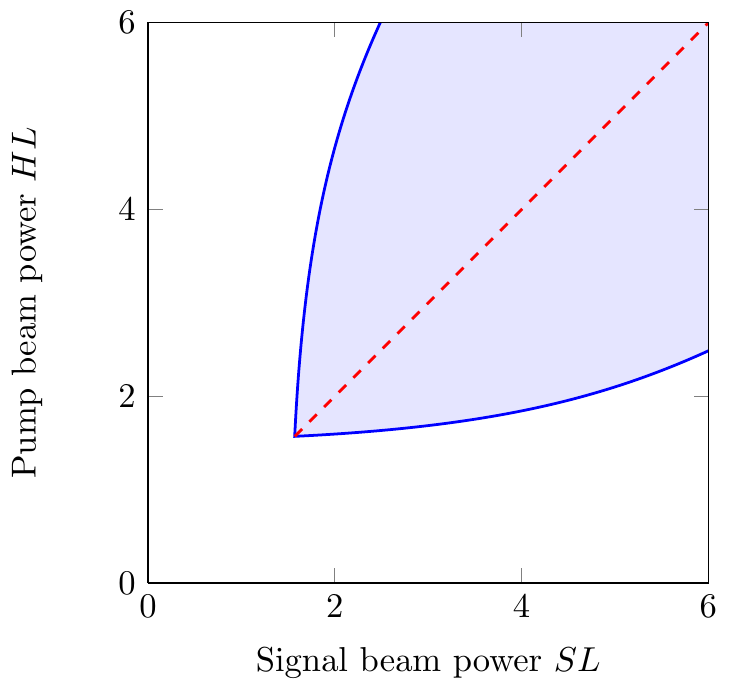}
\caption{Diagram of different polarization attraction regimes. The white region corresponds to the regime of a single branch of stationary solutions: all of these solutions are temporally stable, but the degree of attraction (or output signal-pump alignment) remains relatively low. The shaded region corresponds to the multi-branch regime, where no stationary solutions exist for some values of the input signal-pump relative SOPs. Finally, the dashed red curve shows the most important regime where a stable branch exists for any input signal SOP, and the polarization attraction strength is the highest.}
\label{fig2}
\end{figure}

In order to quantify the strength of the polarization attraction effect, it is useful to analyze the structure of the branches of stationary solutions as in Fig.(\ref{fig-branches}) and Fig.(\ref{fig-new}), and their dependence on the powers of the signal and pump beams $S$ and $H$. We have observed that three different regimes exist. First, whenever the power of either the signal or the pump is relatively small, only one branch of the stationary solution exists which is always stable (e.g., the case with $HL=SL=1$ in Fig.(\ref{fig-branches})). The corresponding region is shown in white color in figure \ref{fig2}. Whenever both the signal and the pump powers are large enough, multiple solution branches appear. For example, consider the case with $HL=5,\!SL=3$ in Fig.(\ref{fig-new}): as can be seen, for some values of the initial signal-pump polarization alignment $\mu$, it turns out that there are no stable stationary solutions at all. A more detailed numerical analysis of these regimes is necessary in order to understand the structure of the non-stationary solutions. However, from a practical perspective this is not a very interesting situation, as one cannot ensure an efficient polarization attraction in this case. The corresponding region is shaded in grey in figure (\ref{fig2}). 

Finally, there is a third regime which corresponds to the line $H=S$ with values $H,S>H_{crit} = \pi/2L$ (red dashed curve in figure (\ref{fig2})). In this regime, the lowest branch (e.g., see the red dashed curve with $HL=SL=2$ in Fig.(\ref{fig-branches})) covers the whole range of initial polarization alignments $-1\leq \mu\leq 1$: correspondingly, the strength of polarization attraction is very high. Clearly this is the most interesting regime from the practical viewpoint, although it might be challenging to achieve since the power of the pump beam needs to be locked to the power of the signal.

\begin{figure}[ht]
\centering
\includegraphics[width=\columnwidth]{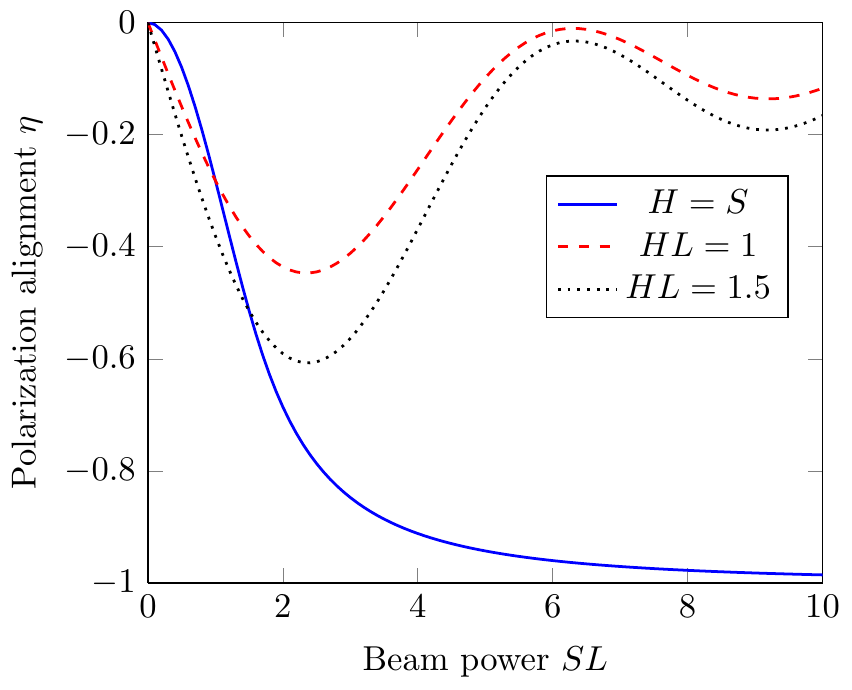}
\caption{Output signal-pump average polarization alignment parameter vs. signal beam power, for different values of pump power. Optimal polarization attraction is observed for relatively high power values, with matched values of signal and beam powers $S = H$.}
\label{fig3}
\end{figure}

As a matter of fact, it is possible to quantitatively characterize the average strength of the polarization attraction process in the situation where the initial signal beam has a random SOP. In this case, the average output signal-pump alignment is simply given by the following expression (we assume that we are operating in the regime where the lowest branch covers the whole region of initial polarization alignments $\mu$):
\begin{equation}
 \langle\eta\rangle = \frac{1}{2}\int_{-1}^1\eta(\mu) d\mu = -\frac{1}{2}\int_{-1}^1 F(\eta) d\eta
\end{equation}
The resulting dependence of the average output signal-pump alignment parameter on the input power of the signal beam is shown in figure (\ref{fig3}). Here we compare the case of equal signal and pump powers or $H=S$ (solid curve) with different situations where their power is unequal (dashed and dotted curves). As it can be seen in Fig.(\ref{fig3}), with matched signal and pump powers there is a monotonic decrease of the average pump-signal alignment parameter $\langle\eta\rangle$ from zero towards $\langle\eta\rangle=-1$ for high beam powers. On the other hand, with unequal signal and pump powers the average alignment $\langle\eta\rangle$ exhibits an oscillating behavior as a function of the signal power, without reaching a significant degree of orthogonality (i.e., $\langle\eta\rangle\cong-1$) even for relatively high signal powers. Thus the results of Fig.(\ref{fig3}) confirm the previous statement that effective polarization attraction only occurs whenever the pump and signal beam power values are located on the red dashed curve in Fig.(\ref{fig2}).    


\section{Stability Analysis}
\label{sec:stability}

\begin{figure}[ht]
\centering
\includegraphics[width=\columnwidth]{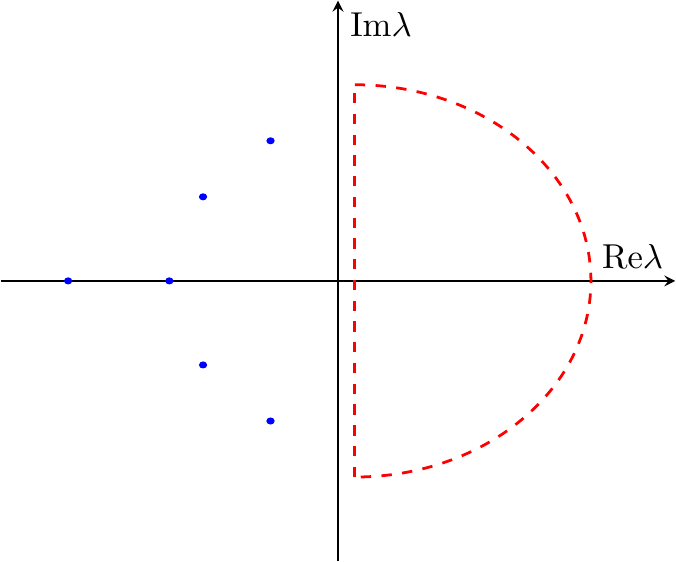}
\caption{Schematic representation of the Nyquist contour used for stability analysis (red dashed curve) and possible locations of eigenvalues of stable dynamic system (blue dots).}
\label{fig-stab}
\end{figure}

The stability of the stationary solutions can be studied with the help of the linearized equations of motion (\ref{main1}):
\begin{eqnarray}
& \left(\partial_t + \partial_z \right) \textbf{s} = \textbf{H}\times\textbf{s} -  \textbf{S}\times \textbf{h} \nonumber \\
\\
& \left(\partial_t - \partial_z \right) \textbf{h} = \textbf{H}\times\textbf{s} -  \textbf{S}\times \textbf{h} \nonumber
\end{eqnarray}

Both $\textbf{H}$ and $\textbf{S}$ depend on the position $z$ according to the expression (\ref{SLdyn}) for the solution found above. However, it is possible to simplify the equations by turning into a reference frame rotating with ${\bf\Omega} = \textbf{H} + \textbf{S}$, where both $\textbf{H}$ and $\textbf{S}$ become constant vectors. Formally this corresponds to the transformation $\textbf{a}(z) \to \exp(\hat{\Omega}z)\textbf{a}(z)$, applied to every vector $\textbf{a} = \textbf{h},\textbf{s},\textbf{H},\textbf{S}$. The operator $\hat{\Omega}$ represents the vector cross product operation: $\hat{\Omega} \textbf{a} = {\bf\Omega}\times\textbf{a}$. This yields the following equations:
\begin{eqnarray}
 &\left(\partial_t + \partial_z \right) \textbf{s} =  -\textbf{S}\times \left(\textbf{h}+\textbf{s}\right), \nonumber \\
\\
 &\left(\partial_t - \partial_z \right) \textbf{h} =  \textbf{H}\times \left(\textbf{h}+\textbf{s}\right). \nonumber
\end{eqnarray}

These equations have to be complemented with the appropriate boundary conditions: $\textbf{s}(z=0,t) = 0$ and $\textbf{h}(z=L,t) = 0$. In order to analyze the stability of small deviations on top of the stationary solutions, we turn to the Laplace transform of $\textbf{s},\textbf{h}$. Assuming that $\textbf{s} = \textbf{s}_\lambda e^{\lambda t}$ and $\textbf{h} = \textbf{h}_\lambda e^{\lambda t}$, we obtain the following system of ordinary differential equations:
\begin{eqnarray}
&\partial_z{ \textbf{s}_\lambda } = - \lambda \textbf{s} - \textbf{S}\times \left(\textbf{h}+\textbf{s}\right), \nonumber \\
\\
&\partial_z{ \textbf{h}_\lambda } = \lambda \textbf{h} - \textbf{H}\times \left(\textbf{h}+\textbf{s}\right). \nonumber
\end{eqnarray}

This system of equations can be written in the matrix form after introduction of the vector $\psi^T = [\textbf{s}^T\,\, \textbf{h}^T]$ and of the matrix form of the cross product operation $\hat S x = \textbf{S}\times x $ and $\hat H x = \textbf{H}\times x $. The resulting equation has the form
\begin{eqnarray}\label{eq:linear}
\partial_z\psi = \hat A \psi = \left [ \begin{array} {cc} -\lambda \hat{1} - \hat S& -\hat S \\ -\hat H & \lambda \hat 1 - \hat H \end{array}\right]\psi,
\end{eqnarray}
where $\hat 1$ is $3\times 3$ identity matrix. The equation (\ref{eq:linear}) is a linear ODE with constant coefficients. The spectrum of the linear normal modes that exist on top of stationary nonlinear solutions can be found by using the boundary condition equations. The solution of the Cauchy problem associated with (\ref{eq:linear}) can be formally written as $\psi(L) = \exp(L \hat A)\psi(0)$. The solution satisfying the boundary conditions exists whenever the solution corresponding to initial conditions $\psi(0) = [0\,\, \textbf{h}_0^T]^T$ that satisfy $\textbf{s}(z=0,t) = 0$ also satisfies the boundary condition $\textbf{h}(z=L,t) = 0$. In other words there exists a solution of the following system of equation: $\hat P  \exp(L \hat A) \hat P^T \textbf{h}_0 = 0$, where the projection operator $\hat P$ is given by $\hat P = [\hat 0\,\,  \hat 1]$ with $\hat 0$ being $3\times 3$ zero matrix. The Wronskian corresponding to this boundary value problem is thus given by
\begin{eqnarray}
 W(\lambda) = \det\left(\hat P  \exp\left[L \hat A(\lambda)\right] \hat P^T\right).
\end{eqnarray}

The eigenmodes of the system correspond to the roots of the Wronskian: $W(\lambda) = 0$ and therefore the stability of the system can be assessed by finding the number of roots in the right side of the complex plane $\mathrm{Re}\lambda > 0$. As the only singularity of the Wronskian function is at $\lambda=\infty$, the number of roots $n_+$ in the right side of the complex plane can be expressed via an integral over the contour surrounding the complex plane according to the classical Cauchy argument principle:
\begin{eqnarray}
 n_+ = \frac{1}{2\pi i}\oint_\Gamma \frac{W'(\lambda) d\lambda}{W(\lambda)}
\end{eqnarray}

The traditional choice of the contour $\Gamma$ attributed to Nyquist is composed of the imaginary axis $\lambda = i y$ with $y \in [-L,L]$ and a half circle $L\exp(i\phi),\,\,\phi \in [-\pi/2,\pi,2]$ with $L\to\infty$.

After the implementation of this procedure with the Mathematica software and its extensive testing in a wide range of parameters, we have found that only the lowest branch of the nonlinear stationary solutions is stable. We are not aware of any analytical proof of this statement, although it should be possible to derive it with an accurate analysis of the algebraic structure of the problem.

\section{Discussion and Conclusions}
\label{sec:concl}

The availability of the analytical expression (\ref{Ffunc}) for the relationship between initial $\mu$ and final $\eta$ polarization alignment between the SOP of the signal and the input pump SOP has permitted us to obtain a relatively simple, and yet general description of the origin of the polarization attraction effect in randomly birefringent telecom optical fibers.
In fact, for observing any polarization attraction it is necessary that the powers of both the signal and the pump are larger than a certain threshold value, so that multiple values of the output alignment $\eta$ result for a given value of the input alignment $\mu$. Moreover, our analysis predicts that the strength of polarization attraction is substantially enhanced whenever the signal and pump beam powers are kept equal. A numerical yet rigorous temporal stability analysis has confirmed numerical simulation results showing that the temporally stable stationary solutions are only those situated on the lower branch of the optical polarization multistability curves such as those reported in Fig.(\ref{fig-branches}) \cite{kozlov_unpub}.

These results provide an interesting insight into the optimal conditions for experimentally observing polarization attraction in long spans of randomly birefringent telecom optical fibers. Therefore we expect that they will find an useful application in fiber-based devices for the all-optical and potentially ultrafast control of the light SOP in optical communication systems as well as in fiber lasers.  

\section{Acknowledgments}

We are especially grateful to Victor V. Kozlov, who has inspired and motivated us during the early stages of this investigation, and who unfortunately could not see the end of this work because of his tragic decease. 
\appendix

\section{Appendix A}
\label{sec:appA}






\end{document}